\documentstyle[epsfig,cite]{mn}

\begin{document}

\title[The orbital modulation in the radio emission of Cygnus~X-1]
  {The orbital modulation in the radio emission of Cygnus~X-1}
\author[Brocksopp, Fender, Pooley]
  {C.~Brocksopp$^1$\thanks{email: cb@astro.livjm.ac.uk}, R.P. Fender$^2$, G.G.~Pooley$^3$\\
$^1$Astrophysics Research Institute, Liverpool John Moores University, Twelve Quays House, Egerton Wharf, Birkenhead CH41 1LD\\
$^2$Astronomical Institute ``Anton Pannekoek'' and Center for High-Energy Astrophysics, University of Amsterdam, \\
\hspace{0.5cm}Kruislaan 403, 1098 SJ Amsterdam, The Netherlands\\
$^3$Mullard Radio Astronomy Observatory, Cavendish Laboratory, Madingley Road, Cambridge CB3 0HE\\}
\date{Accepted ??. Received ??}
\pagerange{\pageref{firstpage}--\pageref{lastpage}}
\pubyear{??}

\maketitle

\begin{abstract}
We present model lightcurves which have been created in order to explain the orbital modulation observed in the radio emission of Cyg X-1. We invoke variable absorption by the stellar wind as the black hole jet orbits around the OB companion star and find that a very simple model is able to reproduce the amplitudes and frequency dependence of the observed lightcurves.

\end{abstract}

\begin{keywords}
stars:individual:Cyg X-1 --- stars:winds, outflows --- radio continuum:stars --- binaries:general
\end{keywords}

\section{Introduction}

Cygnus X-1 was first observed to be a radio source by Braes \& Miley (1971). At the time of discovery the radio flux was $\sim 20$ mJy and it generally maintains a constant mean level of $\sim$ 10--25 mJy at cm wavelengths while the system is in its usual low/hard X-ray spectral state. However, when Cyg X-1 makes one of its rare state changes to the soft state the radio emission appears to be quenched and drops to below a detectable level (Pooley 2001; Brocksopp et al. 1999a, and references within). This behaviour is similar to that of GX 339$-$4 -- another black hole X-ray binary which has been monitored in the radio throughout a soft state period and transitions (Fender et al. 1999, Corbel et al. 2000).

An important feature of the radio emission is the observed flat synchrotron spectrum to millimetre wavelengths (Fender et al. 2000) with no evidence for a cut-off at higher frequencies. Unfortunately the infrared emission is dominated by the supergiant and so it is not possible to determine how far the flat spectrum continues. This spectrum suggests that Cyg X-1 has a quasi-continuous jet of the type modelled by Hjellming \& Johnston (1988) and/or Blandford \& K\"onigl (1979). These predictions have recently been confirmed with the VLBI imaging of the jets of Cyg X-1 (Stirling et al. 2001).

Two types of periodic behaviour can be observed in the radio emission. The longer of these is at $\sim$140 days (although is probably not stable in the long term) and is possibly due to precession of the disc-jet system (Brocksopp et al. 1999a); alternatively a variable mass accretion rate such as that proposed for LMC X-3 (Wilms et al. 2001; Brocksopp, Groot \& Wilms 2001) is also a potential cause of this long period. The 5.6-day orbital period can also be detected in the radio emission (Pooley, Fender \& Brocksopp 1999) and it is this modulation that we now investigate.

The orbital period of Cyg X-1 has been well-established for more than two decades and particularly since the work of Gies \& Bolton (1982). The ephemeris has been revised recently (e.g. Brocksopp et al. 1999b, Sowers et al. 1998, LaSala et al. 1998) but the results have all been consistent with each other. 

X-ray data have also revealed the orbital period (Paciesas et al. 1997) and in Brocksopp et al. (1999a) we discussed the significance of finding the orbital period at all wavelengths in the Cyg X-1 system. Detection of the orbital period of black hole X-ray binaries in optical and infrared photometry is standard and double-peaked orbital lightcurves, due to the gravitational pull on the star by the black hole, are expected -- minima occur in the lightcurves at the two conjunctions. However, it is not so common to detect the orbital period in X-ray or radio data, particularly since Cyg X-1 is a non-eclipsing system and is thought to have a circular orbit. Thus Brocksopp et al. (1999a) postulated that these modulations could be the result of absorption by the stellar wind. This was later confirmed by Wen et al. (1999) who studied the orbital modulation of the $RXTE$ soft X-ray data and produced a relatively successful model of wind absorption. The X-ray orbital lightcurves have also been studied by Ba{\l}uci\'{n}ska--Church et al. (2000) who found that the duration and phase of the minima are also variable and likely to be due to `clumpiness' in the stellar wind.

At radio wavelengths we found that the modulation is frequency-dependent with the strongest modulation at higher frequencies -- contrary to what one would expect for straightforward free-free absorption of a point source. There is also a phase lag present, the duration of which may also be frequency-dependent. These observations add support to the proposed continuous jet model on account of an extended jet-like structure probing different densities of an absorbing medium. Furthermore it would be unlikely that the modulations could be maintained as consistently as observed, were the radio emission produced in discrete ejections.

We now produce a model of wind absorption that can explain the observed radio emission at three wavelengths. To do this we use the focussed stellar wind model of Friend \& Castor (1982), which they and Gies \& Bolton (1986b) have shown to be applicable to the Cyg X-1 system. In this model the spherical stellar wind of the supergiant becomes distorted by the gravitational field of the black hole, thus resulting in enhanced emission occuring close to the axis joining the supergiant and black hole.

%*****************************************************************************
\section{Observations}

We have 15 GHz radio data from the Ryle Telescope of the Mullard Radio Astronomy Observatory, Cambridge, UK; details of the observations can be found in Brocksopp et al. (1999a, and references therein). Further data have been obtained from the Green Bank Interferometer which monitored a number of sources at 2.25 and 8.3 GHz up until October 2000 -- details are given in Waltman et al. (1994).

The data points at each frequency were taken between 1996 November and 1999 June and have had the flaring episodes removed in order to smooth the lightcurves. They were then folded on the orbital period according to the ephemeris of Brocksopp et al. (1999b).

\begin{figure*}
\vspace{-.5cm}
\begin{center}
\leavevmode 
\psfig{file=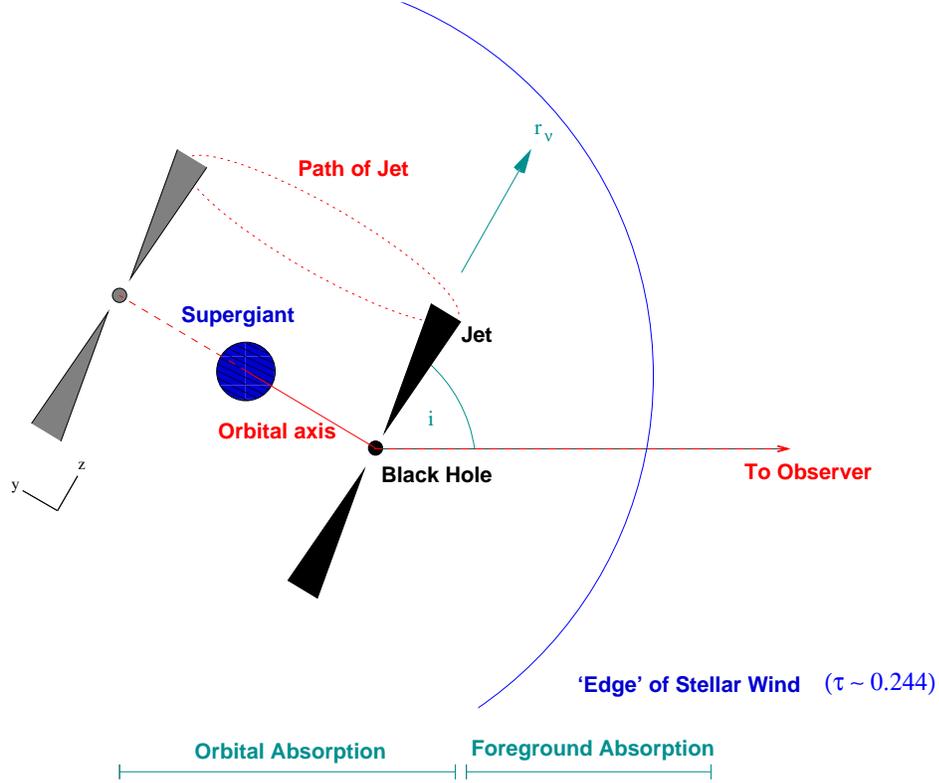, angle=0, width=15cm} 
\vspace{1cm}   
\caption{Schematic showing the Cyg X-1 system as considered in this simple model for the radio modulation. Note that the jets extend considerably further than shown in this diagram. The `edge' of the wind is defined as where the optical depth of the wind (measured from infinity) drops to $\tau\sim0.244$ (Wright \& Barlow 1975).}
\label{model}
\end{center}
\end{figure*}

%*****************************************************************************
\section{Model lightcurves}

In Pooley, Fender \& Brocksopp (1999) it was discovered that the radio emission of Cyg X-1 was modulated on the orbital period with an amplitude that, perhaps surprisingly, increased with the frequency of the emission; not what we would expect for a typical free-free absorption law. This indicates firstly that the density of the wind cannot be treated as constant along the line of sight -- indeed, there is {\em no} stellar wind model which considers the wind density to be constant -- and secondly that the optical depth along the jet is not constant.

We use the work of Leitherer, Chapman \& Koribalski (1995, and references within) to determine the (frequency dependent) radius from the supergiant at which the stellar wind becomes optically thin to free-free absorption of radio emission ($\tau\sim0.244$, measured from infinity; Wright \& Barlow 1975):

\vspace*{-0.25cm}
\begin{equation}
W_{\nu}=4.0\times10^{17}T^{-1/2}g_{\nu}^{1/3}\bigg(\frac{\dot{M}}{v_{\infty}\nu}\bigg)^{2/3}\,\,\,\,\,\,R_{\odot}
\end{equation}

\hspace*{-0.5cm}where we assume a wind temperature $T\sim 5000$K (assuming the results of e.g. White \& Becker 1983 and Leitherer, Chapman \& Koribalski 1995), use $\dot{M}\sim2.0\times10^{-6}$ M$_{\odot}$/yr (Friend \& Castor 1982) and $v_{\infty}\sim 1580$ km/s (Gies \& Bolton 1986b) and calculate the Gaunt factor, $g_{\nu}$ (see Leitherer, Chapman \& Koribalski 1995).

The resultant values of $W_{\nu}$ are $4.6\times10^{14}$, $1.8\times10^{14}$ and $1.2\times10^{14}$ cm for $\nu=$ 2.25, 8.3 and 15 GHz respectively. These values are considerably greater than the binary separation and so we must consider two different absorption components: phase-dependent absorption takes place within the binary orbit, providing the modulation, and  additional (non phase-dependent) absorption takes place in the foreground between the orbit and the wind radius (which lowers the flux by a constant amount).

We assume that the radio emission is absorbed by the wind via the free-free mechanism (Rybicki \& Lightmann 1979):

\vspace*{-0.25cm}
\begin{equation}
\kappa=0.018T^{-3/2}z^2n_en_i\nu^{-2}\bar{g}_{ff}=\frac{3.46\times10^{-19}\nu^{-2}\rho^2}{(m_e+m_i)^2}\,\,\,\,\mbox{cm}^{-1}
\end{equation}

\hspace*{-0.5cm}Then the observed flux is:

%\vspace*{-0.25cm}
\begin{equation}
S_{\nu}=S_0e^{-(\tau_{\phi}+\tau_{f})}=S_0e^{-\int(\kappa_{\phi}+\kappa_{f})dr}
\end{equation}

\hspace*{-0.5cm}where $S_0$ is the intrinsic flux density of the radio source and the absorption components, $\kappa_{\phi}$ (phase-dependent) and $\kappa_{f}$ (foreground) are determined below.

Unfortunately there is no way of knowing what the intrinsic flux density of the radio source is -- therefore we have used the observed flux density of 14.0 mJy $\pm$ 4.0, 6.0, 1.2 mJy for the 2.25, 8.3 and 15 GHz observations respectively (i.e. a flat spectrum); this value is consistent with the peak flux density at each frequency, allowing for the errors of the GBI and Ryle telescopes. The value of the intrinsic flux density is somewhat arbitrary but its variation leads to minimal change in the amplitude of the modulation and so merely has the effect of shifting the model lightcurve up or down.

In order to estimate the distance along the jet at which each frequency becomes optically thin ($\tau\sim1$), we note that the orbital modulation at 2.25 GHz is negligible. Thus we assume that the 2.25 GHz emission site is approximately coincident with the wind radius calculated previously. We then use the relationship $r_{\nu}\propto\nu^{-1}$, as determined by Blandford \& K\"onigl (1979) for a compact flat spectrum jet, to estimate the emission sites of the 8.3 and 15 GHz flux. Thus the emission site is located at:

\begin{equation}
r_{\nu}=4.6\times10^{14}\times\frac{2.25}{\nu/\mbox{GHz}}\,\,\,\mbox{cm}
\end{equation}
\vspace*{0.05cm}

We note that this frequency dependence on distance along the jet has already been observed, the best example of which is GRS 1915+105 (Mirabel et al. 1998).

\subsection{Phase-dependent absorption}

The focussed stellar wind model of Gies \& Bolton (1986b) and Friend \& Castor (1982) is used. Since we are interested in the radio emission (which originates a long way from the orbital plane) we use the parameters derived for the region $>20^{\circ}$ from the binary axis. We take Equation 2 of Gies \& Bolton (1986b) for the wind density:

%\vspace*{-0.25cm}
\begin{equation}
\rho_{\phi}=\bigg(\frac{r^{*}}{R}\bigg)^2\bigg(\frac{\rho_0}{(1-r^{*}/R)^{a}}\bigg)
\end{equation}
\vspace*{0.05cm}

where $r^*$ is the radius of the supergiant ($\sim1.387\times10^{12}$ cm) and $a$ is a constant (=1.05) as defined in Gies \& Bolton (1986b). This is substituted into the equation for $\kappa$ above:

%\vspace*{-0.25cm}
\begin{equation}
\kappa_{\phi}=\frac{6.4\times 10^{60}R^{-4}\nu^{-2}}{(1-1.387\times10^{12}R^{-1})^{2.1}}
\end{equation}

\hspace*{-0.5cm}Thus the phase-dependent absorption can be determined at any point $R(x,y,z)$ where $x=y=z=0$ corresponds to the centre of the OB star.

\subsection{Foreground absorption}

Similarly we determine the foreground absorption for any value of $R$. However, once outside the orbit the equation of Gies \& Bolton (1986b) is no longer valid -- there it is sufficient just to consider the stellar wind as a spherical shell expanding at the value of $v_{\infty}$ as far as $W_{\nu}$.

%\vspace*{-0.25cm}
\begin{equation}
\rho_{f}=\frac{-3\dot{M}}{8\pi v_{\infty}}(R^{-2}-W_{\nu}^{-2})
\end{equation}

\vspace*{0.05cm}

Again, this is substituted into Eq. 2 for $\kappa$ above:

%\vspace*{-0.25cm}
\begin{equation}
\kappa_{f}=2.83\times10^{19}\nu^{-2}(R^{-2}-W_{\nu}^{-2})^2
\end{equation}

The foreground absorption was actually found to be negligible ($\tau_{f}<10^{-36}$ between the `edge' of the wind and the binary orbit) and all significant absorption took place as the black hole passed behind the stellar companion -- this is in agreement with Gies \& Bolton (1986a) who determine that the density of the wind beyond the binary orbit is very small. We note that recombination of the ionised wind at large distances from the supergiant has not been taken into account -- therefore we might have expected to {\em over}-estimate $\kappa_f$.

\subsection{Results}

\begin{figure}
\begin{center}
\leavevmode
\psfig{file=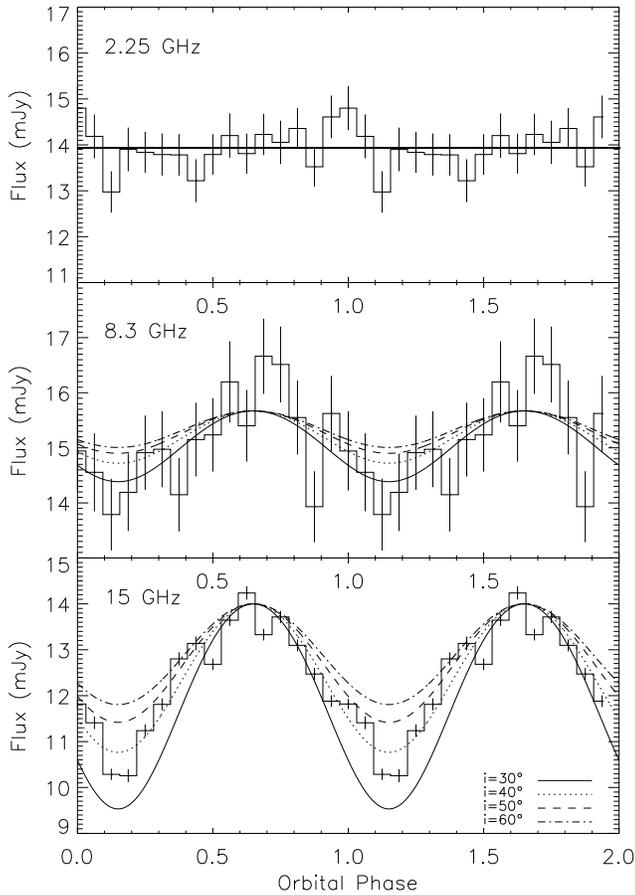, angle=0, width=9cm} 
\vspace{0.25cm}   
\caption{The radio data at 2.25, 8.3 and 15 GHz folded on the orbital period
. Theoretical curves are also plotted showing that the simple model fits the data reasonably well.}
\label{model}
\end{center}
\end{figure}

With knowledge of the absorption at each point in the stellar wind and the position of the radio emission site, it was possible to integrate along the line of sight from the observer to the circle traced by the jet about the star -- i.e. a circle centred on $x=0, y=0, z=r_\nu$. This was repeated for 360 positions around the circle in the frame of the binary system. It was assumed that the angle between the line of sight and the circle traced by the jet remained constant and was equivalent to the inclination angle of the binary system. It was also assumed that all radio emission comes from a single jet -- Doppler boosting of the approaching jet and Doppler de-boosting of the receding jet lead to significant differences between the observed flux of each and so this is indeed a reasonable assumption.

The resultant model lightcurves are shown in Fig.~\ref{model}, plotted over the folded data. Each plot shows four different curves, corresponding to a range of inclination angles. We note that while the curves appear sinusoidal the peaks are in fact slightly broader than a sinusoid -- this is in agreement with previous attempts to fit a sine wave to the mean lightcurves which were somewhat unsuccessful (Pooley, Fender \& Brocksopp 1999). The reduced $\chi^2$ between the data and the model has been determined and the results listed in Table~\ref{chi}.

\begin{table}
\begin{center}
\begin{tabular}{lcccc}
\hline
Frequency&$30^{\circ}$&$40^{\circ}$&$50^{\circ}$&$60^{\circ}$\\
\hline
2.25 GHz&0.45&0.45&0.45&0.45\\
8.3 GHz&4.9&4.2&3.9&3.7\\
15.0 GHz&4.75&1.15&2.40&4.18\\
\hline
\end{tabular}
\end{center}
\caption{Values of the reduced $\chi^2$ between the data and the model at each of the four inclination angles plotted in Fig.~\ref{model}. The model fits reasonably well, particularly at 15 GHz for which the data quality is significantly better.}
\label{chi}
\end{table}

Since the model has been defined to fit the 2.25 GHz data with zero modulation there is not much that can be said about the top plot in Fig.~\ref{model}. However, its lack of orbital variability shows that we were justified in considering the 2.25 GHz emission to be produced outside the wind radius. The 8.3 and 15.0 GHz lightcurves are reasonably well-fit by the model, particularly at an inclination angle of 40$^{\circ}$ which is consistent with values quoted in the literature. Clearly it would be beneficial to improve on the sensitivity of the 2.25 and 8.3 GHz data in order to confirm the quality of the fit. We would, however, like to add a note of caution regarding the use of our work to predict the inclination angle of the system -- the focussed stellar wind model of Gies \& Bolton (1986b), on which our model is largely based, assumes an inclination $\sim30^{\circ}$ and so to suggest that our model predicts $i\sim40^{\circ}$ is not necessarily appropriate without further development of the wind model as suggested below.

\begin{figure}
\begin{center}
\leavevmode
\psfig{file=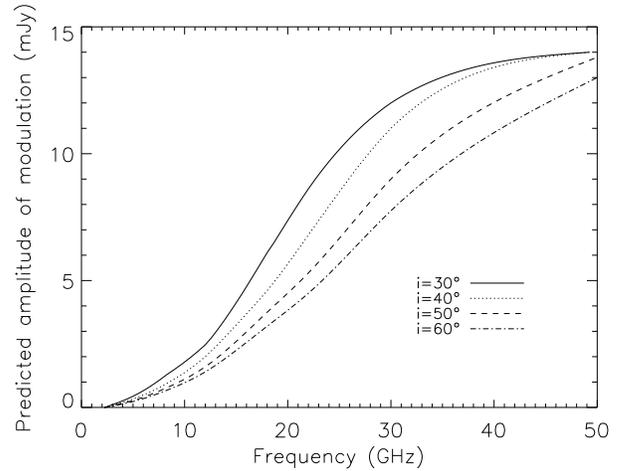, angle=90, width=9cm} 
\vspace{0.25cm}   
\caption{Predicted amplitudes for the orbital modulation at a range of radio frequencies. The model suggests that the emission is saturated at superior conjunction of the black hole for high frequencies and we hope to test this observationally with SCUBA observations.}
\label{amplitude}
\end{center}
\end{figure}

We can also use our model to predict the amplitude of the orbital modulation at radio frequencies other than those observed here. The results are shown in Fig.~\ref{amplitude} for four values of the inclination angle and it is clear that the degree of absorption at superior conjunction of the black hole increases with frequency. Once the frequency has been increased to $\sim$50 GHz the model predicts saturation at an orbital phase of $\sim 0.15$ and this may be expected given that the high frequency emission is produced right at the base of the jet in the most dense regions of the stellar wind -- this is yet to be tested observationally and we have proposed millimetre observations with which to do this.

%************************************************************************

\section{Discussion}
\subsection{Improving the model}

The most obvious problem with this basic model is that it has been assumed that the radio emission at each frequency is essentially a point source at position $r_{\nu}$ along the jet, and that this position lies within the stellar wind. Since this assumption results in reasonable theoretical curves it is tempting to consider this to be the case. However the VLBA maps of Stirling et al. (2001) show that, while it is a reasonable assumption for the 15 GHz emission, it is unlikely to be the case at 8.3 GHz -- at each epoch there is a significant fraction of the jet well outside the wind radius and it is clearly not a point source. Re-running the code to allow for $\sim50\%$ of the emission to be outside the modulated region produces too little modulation compared with the GBI data-set. Certainly it would be beneficial to obtain higher sensitivity data in order to check this discrepancy. Alternatively it is possible that the wind density in the outer parts of the wind has been underestimated and that the foreground absorption is actually significant -- a higher intrinsic flux density would allow for $<100\%$ of the emission to be modulated and to still produce sufficiently deep minima. Problems with the wind model are discussed below.

There are a number of further assumptions which have been made in order to create the model lightcurves -- we have not taken account of the wind structure, other than the simple assumption that we can use the focussed wind model as described above. Firstly it is unlikely that the wind is a constant temperature throughout -- with the stellar surface thought to be at a temperature of $\sim 30000$ K there will be a region of hotter and less dense wind near the star. This may not affect the radio emission (at least at lower frequencies) since it comes from a region a long way from the star. Secondly we have assumed a constant density function throughout the wind. Again this is unlikely as the wind will be clumpy and irregular. Adding these corrections to the model at this stage is well beyond the scope of this paper -- detailed modelling of the stellar atmosphere and wind in the Cyg X-1 system is planned but is a long-term project.

A further effect on the stellar wind is caused by the presence of the X-ray source. The stellar wind of an OB star, such as the optical counterpart to Cyg X-1, is emitted by radiation pressure in the star's outer atmosphere which scatters and absorbs photons in various spectral lines formed in the wind. The acceleration of the wind is influenced by the degree of X-ray heating and ionisation from a compact object situated in the path of the wind (Hatchett \& McCray 1977). While an increase in X-ray flux increases the line-driving force and hence the wind velocity, once the X-ray luminosity is of the order $10^{34}$ ergs/s the line force and wind velocity begin to decrease again. This is due to the increasing proportion of highly ionised species (emitting at optical, ultraviolet and X-ray wavelengths), which leads to a decrease in luminosity of certain emission lines and subsequent alteration of the wind spectrum\footnote{An effect of this appears to be observed -- when the X-ray source of Cyg X-1 makes a transition to the high/soft state (i.e. an increased source of soft X-ray photons at the centre of the accretion disc), a quenching of the H$\alpha$ focussed wind emission is detected via equivalent width measurements (Voloshina, Lyuty \& Tarasov 1997). We note that the radio jet is associated with the low/hard state only (Brocksopp et al. 1999a) and so this observation is not relevant to the model presented here.} (Hatchett, Buff \& McCray 1976). By the time the X-ray luminosity reaches $5\times 10^{34}$ ergs/s the high ionisation in the region around the compact object is sufficient to overcome the line-driving force, thus quenching the radiative acceleration in a region known as the Str\"omgren zone (Macgregor \& Vitello 1982). With a luminosity of $\sim 10^{37}$ erg/s, the X-ray source of Cyg X-1 should produce a significant effect on the wind. However, assuming that the X-ray flux decreases as $\sim 1/r^2$ and that the X-ray ionisation is only significant in the high velocity wind close to the orbital axis (Hatchett, Buff \& McCray 1976), it is probable that this Hatchett-McCray effect is insignificant at the radius of the radio emission sites. (See van Loon, Kaper \& Hammerschlag-Hensberge 2001 for detailed modelling and size estimates of the Str\"omgren zone.)

\subsection{Further considerations}

An additional property that is not represented by the model is the phase offset seen in all radio lightcurves -- the theoretical curves have been shifted in phase by 0.15 in order to match the observations. While Pooley, Fender \& Brocksopp (1999) and Brocksopp et al. (1999b) find a frequency dependence to this offset, it is not confirmed in the extended datasets used here. (The reason for the disppearance of the frequency-dependence to this lag is uncertain -- it is of course possible that it was some artefact caused by the high noise levels in the GBI data. Higher sensitivity VLA observations are planned in order to investigate this effect further.) However, the offset is still very much present in the mean lightcurves. A possible cause of this could be some sort of drag effect -- as the black hole passes through the stellar wind the jets may be `swept back' and trail behind slightly. Frequencies emitted towards the far end of the jet (i.e. lower frequencies) would then be delayed by a greater amount than frequencies emitted nearer the black hole.

Such a phase offset can be simply modelled -- assuming that the phase delay is dominated by self-absorption of the jet, the jet will lag the black hole by a distance $$\frac{r_{\nu}}{v_{jet}}\times v_{orb}$$ where $v_{jet}$ and $v_{orb}$ are the jet and orbital velocities respectively. This corresponds to a phase lag of $$\frac{r_{\nu}}{v_{jet}\times P_{orb}}$$ where $P_{orb}$ is the orbital period. By assuming $r_{15\mbox{\tiny{GHz}}}\sim 6.8\times10^{13}$ cm then a phase lag of 0.15 could be produced by a jet travelling at $v_{jet}\sim 0.03$c -- while this is mildly relativistic, it is probably an underestimate, particularly given the lower limits estimated by Stirling et al. (2001).

As mentioned previously, the presence of the orbital period in the radio emission is unexpected -- indeed the only other systems for which this has been the case are LSI +61$^{\circ}$303 and Cir X-1 (Taylor \& Gregory 1982; Haynes, Lerche \& Murdin 1980). In both cases the period is much longer than that of Cyg X-1 (26.5 and 16.6 days respectively) and the nature of the radio emission is also very different. Rather than the smooth sinusoid of Cyg X-1, these two sources show periodic radio outbursts thought to be produced at periastron of an eccentric orbit. Therefore even if the companion stars emit strong stellar winds (as certainly is the case for LSI +61$^{\circ}$303), it is unlikely that the wind absorption model can be applied here -- it would predict a {\em minimum} at periastron as this is the point at which the wind is most dense. Alternatively the wind model may still be relevant but the effects are dominated by the higher accretion rate at periastron and subsequently greater jet power. However, it is interesting to note that the only previous attempt to model the radio modulation of Cyg X-1 is Han (1993). She also invokes an elliptical orbit with an inclined jet -- this can be discounted given the lack of evidence for anything other than a circular orbit in optical spectroscopy presented in the literature (e.g. Brocksopp et al. 1999b).

One important application and test of this absorption model will be the lightcurve prediction of jet emission at other frequencies, in particular the millimetre and infrared emission close to the black hole. Cyg X-1 has been detected at  millimetre wavelengths but there is insufficient data available to detect any orbital modulation. A comparison of the model and lightcurves at higher frequencies would also be useful in probing the regions of higher ionisation around the black hole.

%*************************************************************************
\section{Conclusions}

We have presented a model to explain the orbital modulation observed in the radio emission of Cyg X-1. By invoking absorption of the radio emission by the stellar wind of the companion star we are able to produce model lightcurves with amplitudes and frequency dependence comparable with the real data. We acknowledge that the fits could be improved with the use of a more sophisticated stellar wind model -- the production of such a wind model is work in progress.

%*****************************************************************************
\section*{acknowledgements}
The Ryle Telescope is supported by PPARC. The Green Bank Interferometer was operated by the National Radio Astronomy Observatory for the U.S. Naval Observatory and the Naval Research laboratory during the time-period of these observations. Many thanks to Roger Hutchings for his help with writing the code and to Simon Clark and John Porter for useful comments.

\bibliographystyle{mn}

\bsp 

\label{lastpage}

\end{document}